\documentclass[twoside,twocolumn,9pt]{article}
\usepackage{extsizes}
\usepackage[super,sort&compress,comma]{natbib} 
\usepackage[version=3]{mhchem}
\usepackage[left=1.5cm, right=1.5cm, top=1.785cm, bottom=2.0cm]{geometry}
\usepackage{balance}
\usepackage{mathptmx}
\usepackage{sectsty}
\usepackage{graphicx} 
\usepackage{lastpage}
\usepackage{placeins}
\usepackage{rotating}
\usepackage{graphicx}
\usepackage{makecell}
\usepackage{multirow}
\usepackage[format=plain,justification=justified,singlelinecheck=false,font={stretch=1.125,small,sf},labelfont=bf,labelsep=space]{caption}
\usepackage{float}
\usepackage{fancyhdr}
\usepackage{fnpos}
\usepackage[english]{babel}
\addto{\captionsenglish}{%
  
}
\usepackage{array}
\usepackage{droidsans}
\usepackage{charter}
\usepackage[T1]{fontenc}
\usepackage[usenames,dvipsnames]{xcolor}
\usepackage{setspace}
\usepackage[compact]{titlesec}
\usepackage{hyperref}
\usepackage{multirow}

\usepackage{cleveref}
\usepackage{epstopdf}

\definecolor{cream}{RGB}{222,217,201}

\begin{document}

\pagestyle{fancy}
\thispagestyle{plain}
\fancypagestyle{plain}{
\renewcommand{\headrulewidth}{0pt}
}

\makeFNbottom
\makeatletter
\renewcommand\LARGE{\@setfontsize\LARGE{15pt}{17}}
\renewcommand\Large{\@setfontsize\Large{12pt}{14}}
\renewcommand\large{\@setfontsize\large{10pt}{12}}
\renewcommand\footnotesize{\@setfontsize\footnotesize{7pt}{10}}
\renewcommand\scriptsize{\@setfontsize\scriptsize{7pt}{7}}
\makeatother

\renewcommand{\thefootnote}{\fnsymbol{footnote}}
\renewcommand\footnoterule{\vspace*{1pt}%
\color{cream}\hrule width 3.5in height 0.4pt \color{black} \vspace*{5pt}} 
\setcounter{secnumdepth}{5}

\makeatletter 
\renewcommand\@biblabel[1]{#1}            
\renewcommand\@makefntext[1]%
{\noindent\makebox[0pt][r]{\@thefnmark\,}#1}
\makeatother 

\renewcommand{\figurename}{\small{Fig.}~}
\sectionfont{\sffamily\Large}
\subsectionfont{\normalsize}
\subsubsectionfont{\bf}
\setstretch{1.125}
\setlength{\skip\footins}{0.8cm}
\setlength{\footnotesep}{0.25cm}
\setlength{\jot}{10pt}
\titlespacing*{\section}{0pt}{4pt}{4pt}
\titlespacing*{\subsection}{0pt}{15pt}{1pt}

\fancyfoot{}
\fancyfoot[LO,RE]{\vspace{-7.1pt}\includegraphics[height=9pt]{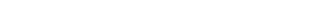}}
\fancyfoot[CO]{\vspace{-7.1pt}\hspace{13.2cm}\includegraphics{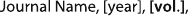}}
\fancyfoot[CE]{\vspace{-7.2pt}\hspace{-14.2cm}\includegraphics{head_foot/RF}}
\fancyfoot[RO]{\footnotesize{\sffamily{1--\pageref{LastPage} ~\textbar  \hspace{2pt}\thepage}}}
\fancyfoot[LE]{\footnotesize{\sffamily{\thepage~\textbar\hspace{3.45cm} 1--\pageref{LastPage}}}}
\fancyhead{}
\renewcommand{\headrulewidth}{0pt} 
\renewcommand{\footrulewidth}{0pt}
\setlength{\arrayrulewidth}{1pt}
\setlength{\columnsep}{6.5mm}
\setlength\bibsep{1pt}

\makeatletter 
\newlength{\figrulesep} 
\setlength{\figrulesep}{0.5\textfloatsep} 

\newcommand{\topfigrule}{\vspace*{-1pt}%
\noindent{\color{cream}\rule[-\figrulesep]{\columnwidth}{1.5pt}} }

\newcommand{\botfigrule}{\vspace*{-2pt}%
\noindent{\color{cream}\rule[\figrulesep]{\columnwidth}{1.5pt}} }

\newcommand{\dblfigrule}{\vspace*{-1pt}%
\noindent{\color{cream}\rule[-\figrulesep]{\textwidth}{1.5pt}} }

\makeatother

\twocolumn[
  \begin{@twocolumnfalse}
{\includegraphics[height=30pt]{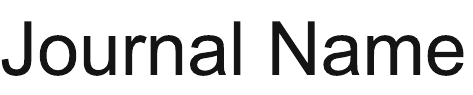}\hfill\raisebox{0pt}[0pt][0pt]{\includegraphics[height=55pt]{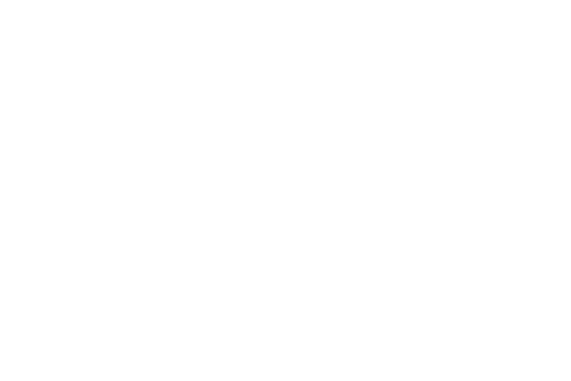}}\\[1ex]
\includegraphics[width=18.5cm]{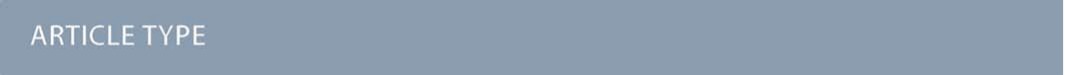}}\par
\vspace{1em}
\sffamily
\begin{tabular}{m{4.5cm} p{12.5cm} }

\includegraphics{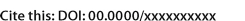} 
& \noindent\LARGE{\textbf{Planar Structures of Medium-Sized Gold Clusters Become Ground States upon Ionization}} \\

& \vspace{0.3cm} \\

& \noindent\large{Mohammad Ismaeil Safa\textit{$^{a}$},
Ehsan Rahmatizad Khajehpasha\textit{$^{a}$}
and Stefan Goedecker\textit{$^{a}$}}$^{\ast}$ \\


\renewcommand*\rmdefault{bch}\normalfont\upshape
\rmfamily
\section*{}
\vspace{-1cm}


\includegraphics{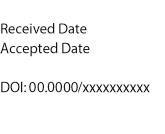} & \noindent\normalsize{This study investigates the structural stability of ionized gold clusters of sizes ranging from 22 to 100 atoms, contrasting compact, cage and planar structures. While it is well known that neutral clusters in the upper part of this size range predominantly favor compact structures, our results reveal that positively ionized gold clusters exhibit structural transitions in which planar structures become energetically preferred once the charge is sufficiently large. In addition, we study the finite-temperature stability of the structures and find that thermodynamic effects further stabilize planar configurations relative to their compact counterparts. To explore the potential energy surface, we use the Minima Hopping algorithm combined with a machine-learned potential. Since the machine-learned potential does not apply to ionized clusters, we introduce a charge-correction term to incorporate Coulomb interactions and charge screening.}

 \end{tabular}

 \end{@twocolumnfalse} \vspace{0.05cm}
 ]


\footnotetext{\textit{$^{a}$~Department of Physics, University of Basel, Klingelbergstrasse 82, CH-4056 Basel, Switzerland.}}

\footnotetext{Corresponding author E-mail: stefan.goedecker@unibas.ch}

\footnotetext{Electronic supplementary information (ESI) available. See DOI: XX.XXXX/XXXXXXXX.}


\rmfamily



\section{Introduction}
Extensive experimental and theoretical studies show that small gold clusters exhibit a preference for planar (2D) geometries up to a critical size, beyond which three-dimensional (3D) compact structures, (i.e structures which can not host an extra endohedral atom), become energetically favorable~\cite{furche2002structures,hakkinen2000gold,hakkinen2003photoelectron,PhysRevB.66.035418}. For anionic clusters, photoelectron spectroscopy, ion mobility measurements and Density Functional Theory (DFT) calculations consistently show that clusters remain planar up to $Au_{11}$, with a structural transition occurring at $Au_{12}$ where planar and 3D isomers coexist and clusters of $Au_{n\geq 13}$ are 3D geometries~\cite{furche2002structures,xing2006structural,pal2011structure,hakkinen2003photoelectron}. DFT predictions for the 2D to 3D crossover in neutral clusters are strongly method-dependent: Local Density Approximation (LDA) favors a transition at $Au_7$, whereas Generalized Gradient Approximation (GGA) with scalar-relativistic can stabilize planar motifs up to $Au_{11}$~\cite{fernandez2006planar}. Cage motifs emerge for $Au_{10}$–$Au_{14}$, followed by a transition to compact structures at $Au_{15}$~\cite{PhysRevB.66.035418}.
Finite-temperature free-energy calculations using van der Waals-corrected DFT predict that $Au_{9,10}$ remain planar at 100 K, whereas $Au_{11}$ adopts a nonplanar capped trigonal prism with $D_{3h}$ symmetry; increasing temperature generally shifts populations toward nonplanar structures for $Au_8-Au_{13}$, with $Au_{11}$ being a notable exception due to many accessible near-degenerate planar isomers~\cite{goldsmith2020two}.
A recent machine-learning-based global search has further refined this picture by systematically investigating $Au_2-Au_{55}$ clusters, identifying the planar to 3D transition at $Au_{14}$ and a subsequent cage to core–shell transformation for $Au_{n\geq26}$~\cite{ding2025revisiting}. 
These results not only confirm earlier DFT-based trends but also provide a comprehensive, size-resolved structural map, including several previously unreported low-energy isomers for $Au_{13}$ and $Au_{15}-Au_{18}$. Additionally, photoelectron spectroscopy and global search studies revealed that anionic clusters $Au_{16}^- - Au_{18}^-$ can also form experimentally detectable cages~\cite{bulusu2006evidence}. A paradigmatic example is $Au_{20}$, which adopts a tetrahedral structure, experimentally confirmed via photoelectron spectroscopy, shown to be a slightly distorted fragment of the bulk fcc lattice with remarkable stability due to its large HOMO-LUMO gap and high electron affinity~\cite{li2003au20}. For $Au_{26}$ and $Au_{26}^{-}$, global searches reveal a fluxional character. Many distinct minima are only  $\sim 0.5$~eV above the putative ground state and are separated by low barriers. In addition, the predicted ground state motif depends on the exchange-correlation (XC) functional: LDA favors filled-cage core-shell structures for $Au_{26}$ and $Au_{26}^{-}$, whereas Perdew–Burke–Ernzerhof (PBE)~\cite{perdew1996generalized}. stabilizes an empty cage for $Au_{26}$ and a tubular motif for $Au_{26}^{-}$~\cite{schaefer2014isomerism}. In line with this picture, high-level coupled-cluster benchmarks for $Au_{27}^q$ already reveal an extremely shallow Potential Energy Surface (PES) with many quasi-degenerate compact and non-compact isomers and a strong XC functional dependence of the predicted ground state~\cite{Au27}. For $Au_{32}$, a cage with icosahedral symmetry satisfying the $2(N+1)^2$ spherical aromaticity rule~\cite{hirsch2000spherical} and exhibiting a large HOMO–LUMO gap is proposed as a “golden fullerene”~\cite{Au32}. Stability, arising from aromaticity and a large HOMO-LUMO gap, was also found for $Au_{50}$, where a DFT-based comparison of cage and compact structures found that a cage with $D_{6d}$ was the global minimum~\cite{tian2006dual, wang2005hollow}. $Au_{42}$ with icosahedral and $Au_{60}$ with chiral icosahedral symmetry cages are typically slightly higher in energy than compacts, remain structurally robust local minima and can act as components in nested or multi-shell clusters~\cite{gao2005au42,mullins2021robustness}. A study based on icosahedral-inspired templates showed that large quasi-icosahedral cages such as $Au_{92}$ and $Au_{122}$, although not global minima, are low-lying energy metastable structures with high symmetry and strong metallicity~\cite{ning2014series}. Clusters in the $Au_{n\sim100}$ size range are compact, and entirely different structural motifs-such as icosahedral, decahedral, or octahedral-can be nearly degenerate in energy. As a consequence, the ground-state structural motif can change upon the addition of a single atom, and the ground-state structure oscillates among these motifs with increasing cluster size~\cite{bao}.

While early studies primarily focused on small clusters and their transition from planar to compact geometries, recent experimental studies have demonstrated that atomically thin, free-standing gold monolayers can be synthesized using ligand-assisted self-assembly~\cite{song2020macroscopic}, intercalation beneath graphene~\cite{forti2020semiconductor}, exfoliation from layered precursors~\cite{kashiwaya2024synthesis} and in-situ dealloying inside electron microscopes~\cite{wang2019free,zhao2020situ}. These monolayers exhibit remarkable thermal stability, metallic conductivity, and even magnetic edge states in nanoribbons. Moreover, studies based on DFT predicted that hexagonally close-packed 2D gold is dynamically~\cite{ono2020dynamical} and thermodynamically~\cite{yang2015glitter} stable. Systematic DFT calculations indicate that graphene pores can stabilize free-standing 2D metal patches up to $\sim 8~{nm}^2$, and identify Au as one of the most promising elemental metals for forming stable pore-confined 2D phases~\cite{nevalaita2019stability}. Complementary computational simulations using ab-initio and molecular dynamics methods further reveal unique behaviors such as 2D liquid phases~\cite{koskinen2015plenty}, strain-induced electronic transitions~\cite{forti2020semiconductor}, and strong environment-dependent catalytic activity~\cite{ye2019sub,marangoni2018engineering,wang2015two}. These findings establish that planar gold structures are not restricted to small clusters but constitute a broader and experimentally realizable class of 2D materials.

These results collectively suggest that planar and cage motifs represent an important class of stable and potentially synthesizable configurations for medium-sized and large gold clusters. We therefore explored whether positive ionization, hereafter referred to simply as ionization, of electron-rich gold clusters provides a route to stabilizing planar or cage motifs. The rationale is based on the assumption that the positive charge is distributed over all atoms, leading to a repulsion between the ionic nuclei. Hence, the electrostatic energy will be smaller in a planar or cage structure, where the number of nearest neighbors is smaller than in compact structures. To test this hypothesis, we systematically investigated the PES of ionized gold clusters ranging from $Au_{22}$ to $Au_{100}$. For this purpose, we employed the Minima Hopping (MH) algorithm~\cite{goedecker2004minima}.

To evaluate the energies and forces during the MH exploration, we used two independent machine learned potentials, the Neural Equivariant Interatomic Potential (NequIP)~\cite{batzner20223} and the Machine-learning Atomic Cluster Expansion (MACE)~\cite{batatia2023foundation}. However, since NequIP is originally designed for neutral systems, we added a physically motivated charge correction term to it. This correction accounts for Columbic repulsion and charge screening effects, thereby enabling reliable modeling of ionized gold clusters.

Our approach uses the Atomic Simulation Environment (ASE)-integrated implementation of MH~\cite{krummenacher2024performing} and Symmetry-biased MH~\cite{huber2023targeting}. Notably, our results reveal that planar geometries emerge as ground state configurations for certain ionization levels, suggesting that ionization can play a pivotal role in stabilizing non-compact cluster morphologies. Furthermore, we account for vibrational contributions to the free energy at finite temperatures. Since atomic vibrations are always present at non-zero temperatures, the vibrational free energy term can affect structural stability. Our calculations indicate that the vibrational free energy contribution at finite temperature further stabilizes planar geometries over compact counterparts. This suggests that, in addition to electrostatic effects, thermal fluctuations also promote the emergence of planar structures at room temperature. 

\section{Methods}

\subsection{Global structure search with MH and machine-learned potentials}

The exploration of the PES for neutral and ionized gold clusters was carried out with the MH algorithm. MH alternates short molecular-dynamics escape moves and local geometry relaxations, while an adaptive temperature scheme controls the kinetic energy used in the escape steps, which makes it well-suited for sampling complex, high-dimensional PES and locating low-energy minima while avoiding excessive revisits of already known structures. In all MH simulations, the interatomic forces and energies were calculated by machine-learned interatomic potentials rather than by DFT, allowing us to perform extensive global searches for cluster sizes up to $Au_{100}$.

As a primary PES model, we employed NequIP, an $E(3)$-equivariant neural network potential that we trained on DFT reference data for gold clusters. With this potential, we could find all previously known neutral cage structures. By construction, this potential does not explicitly include electrostatic contributions associated with ionization. To extend its applicability to ionized clusters, we augmented NequIP with an electrostatic correction term that is added to both the total energy and the atomic forces.

We assumed that, upon ionization, the charge is uniformly distributed over all atomic sites, such that each nucleus carries an effective charge $Q_{tot}/n$, where $Q_{tot}$ is the total cluster charge and $n$ is the cluster size. In our first attempt, we added a bare Coulomb interaction between these point charges on top of the NequIP outputs, but this unscreened correction overestimated the effect of ionization and did not reproduce the DFT trends for ionized clusters. We therefore introduced a screened interaction with an exponential damping factor to mimic charge screening and finite-size effects in the metallic cluster. The resulting energy correction term in atomic units takes the form:
\begin{equation}
    E = \frac{1}{2} \sum_i \sum_j 
     \frac{Q_{\mathrm{tot}}^2}{n r_{ij}} 
    \exp\left(-\frac{r_{ij}}{\lambda}\right),
\end{equation}
where $r_{ij}$ is the distance between atoms $i$ and $j$, and $\lambda$ is a decay length that controls the range of the screened interaction. The corresponding interatomic forces are then the negative gradient of this energy. The exponential damping takes into account screening effects, leading to a smoother and more physical description of nuclear repulsion inside the cluster. In all MH simulations, these correction terms were evaluated on-the-fly and added to the underlying NequIP energies and forces, so that the search is effectively performed on a charge-corrected PES.

Alongside the NequIP-based PES, we employed a MACE foundation model as an independent machine-learned potential. We performed separate MH simulations on the MACE to obtain an alternative global exploration of the low-energy PES. The low-energy structures obtained with NequIP and MACE were subsequently validated with DFT, providing evidence that the emergence of cage and planar ground states is not an artefact of machine-learned potentials.

\subsection{DFT Calculations}

A large number of low-energy compact, cage and planar structures from MH outputs were selected to be validated by DFT. To accurately evaluate the energies and structural properties of ionized gold clusters, we performed DFT calculations using the FHI-aims package \cite{blum2009ab}. FHI-aims employs an all-electron, numeric atom-centered Gaussian basis set, allowing for highly accurate electronic structure calculations. Relativistic effects are treated by the AZORA method~\cite{vanLenthe1993}.

We employed three XC functionals to validate the robustness of the structural trends: PBE, its solid-state revision, PBEsol, combined with a many-body dispersion correction (PBEsol+MBD), both within the GGA framework; and r$^2$SCAN within the meta-GGA framework.~\cite{perdew1996generalized,perdew2008restoring,tkatchenko2012accurate,furness2020accurate}

We employed the "tight" basis set to ensure high numerical accuracy in describing the electronic structure and treated all clusters using free boundary conditions, simulating isolated gas-phase systems without periodic constraints. Performing structural relaxations for neutral clusters and ionization levels up to $+13e$ enabled us to analyze how increasing ionization drives structural transitions. We set the structural-optimization convergence threshold such that the force norm on each atom falls below 5 meV/Å. These calculations provide reference data for benchmarking the machine-learned potentials and analyzing charge-dependent structural transitions in gold nanoclusters.

\section{Results}
In this study, we examined how the structures of gold clusters evolve with increasing ionization. As the ionization level increases, gold clusters show a tendency to form planar structures either in the form of cages or of planar flakes. For selected $Au_n$ clusters with $n\leq 100$ atoms, we successively increased the ionization level until we could observe a transition of the ground state structure into a cage or flake. While a tendency to form planar structures upon ionization is universal, different XC functionals predict different threshold ionizations for this transition. The results for the three widely used XC functionals we employed are presented below.

\subsection{PBE}
\begin{figure*}[htbp]
    \centering
    \includegraphics[width=0.98\textwidth]{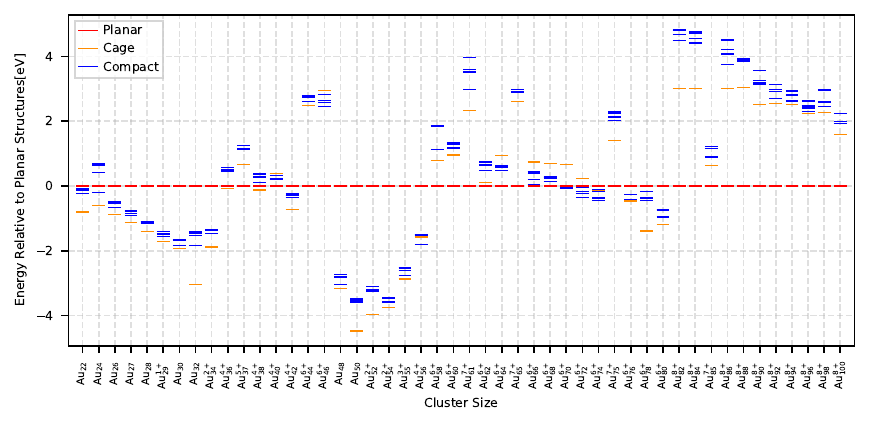}
    \caption{Total PBE energies for the smallest ionization level that gives a cage or flake as its ground state for the PBE XC functional.
    }
    \label{fig:energy_gaps_pbe}
\end{figure*}
We employed PBE, a widely used non-empirical semilocal GGA XC functional. Among the three XC functionals considered in this work, PBE requires the lowest ionization level to stabilize non-compact motifs as ground states. In the neutral clusters, cage geometries are obtained as ground states for $Au_n$ with $n=22,24,26,27,28,30,32,48,$ and $50$. Upon weak ionization, cages become the ground state for $Au_{29}^{+1}$, $Au_{34}^{+2}$, $Au_{36}^{+4}$, $Au_{38}^{+4}$, $Au_{42}^{+4}$, $Au_{52}^{+4}$, $Au_{54}^{+2}$, $Au_{55}^{+3}$, and $Au_{56}^{+4}$. At intermediate ionization level, cage ground states are found for $Au_{70}^{+6}$, $Au_{72}^{+6}$, $Au_{74}^{+6}$, $Au_{76}^{+6}$, $Au_{78}^{+6}$, and $Au_{80}^{+6}$. All remaining clusters favor planar flakes as their ground state structures at the corresponding ionization levels. For the planar ground states, the energetic gaps from the lowest metastable isomers are as follows:
$Au_{82}^{+8}$, $Au_{84}^{+8}$, $Au_{86}^{+8}$ have gaps between $3$-$4$ eV; $Au_{44}^{+6}$, $Au_{46}^{+6}$, $Au_{61}^{+7}$, $Au_{65}^{+7}$, $Au_{90}^{+8}$, $Au_{92}^{+8}$, $Au_{94}^{+8}$, $Au_{96}^{+8}$ and $Au_{98}^{+8}$ have gaps between $2$-$3$ eV; $Au_{100}^{+8}$ and $Au_{75}^{+7}$ have gaps between $1$-$2$ eV; all remaining planar ground states have gaps of less than $1$ eV to the closest metastable structure. The detailed PBE energies and ground state assignments as a function of size and ionization level are shown in~\cref{fig:energy_gaps_pbe}.

\subsection{PBEsol+MBD}
\begin{figure*}[htbp]
    \centering
    \includegraphics[width=0.98\textwidth]{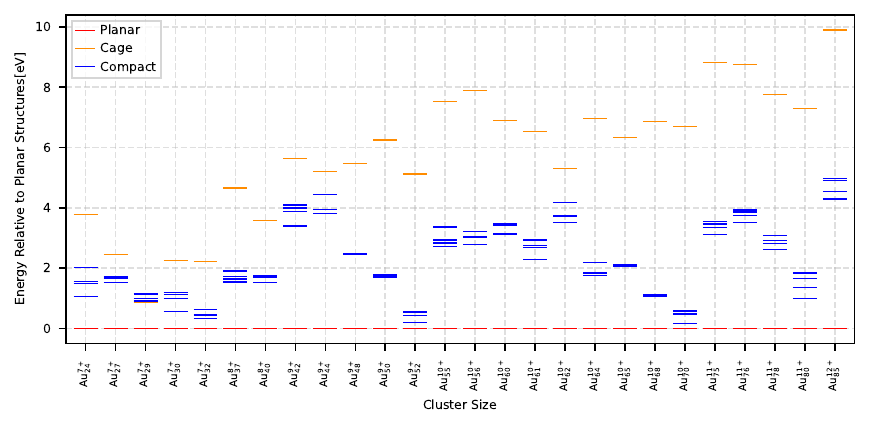}
    \caption{Total PBEsol+MBD energies for the investigated neutral and ionized gold clusters.
    Shown is always the smallest ionization level that gives a cage or flake as its ground state. In contrast to experimental evidence, cages are never lower in energy than planar structures with this XC functional.
    }
    \label{fig:energy_gaps_pbesol}
\end{figure*}
PBEsol is a semi-local GGA XC functional tailored for solids, and the MBD adds an explicit 
many-body dispersion correction. We performed separate relaxations of the same structural candidates to determine, for each size, the smallest ionization level at which a planar structure becomes the global minimum. The ionization level required for planar flakes to become ground states is generally higher than that predicted by the other two XC functionals, while the corresponding energy gaps between the ground state and the first metastable isomer are as follows:
$Au_{85}^{+12}$ with $\Delta E = 4.29$ eV has the largest gap; $Au_{42}^{+9}$, $Au_{44}^{+9}$, $Au_{60}^{+10}$, $Au_{62}^{+10}$, $Au_{75}^{+11}$ and $Au_{76}^{+11}$ have gaps between $3$-$4$ eV; $Au_{48}^{+9}$, $Au_{55}^{10}$, $Au_{56}^{+10}$, $Au_{61}^{+10}$, $Au_{65}^{+10}$ and $Au_{78}^{+11}$ have gaps between $2$-$3$ eV; $Au_{24}^{+7}$, $Au_{27}^{+7}$, $Au_{37}^{+8}$, $Au_{40}^{+8}$, $Au_{50}^{+9}$, $Au_{64}^{+10}$ and $Au_{68}^{+10}$ have gaps between $1$-$2$ eV; all remaining planar ground states have gaps of less than $1$ eV to the closest metastable structure. This XC functional does not favor cage motifs at all; whenever the planar flakes become ground state, the cages are higher in energy than compacts. In the~\cref{tab:compact_symmetries}, the ionization levels required for each cluster size are reported. The resulting energetic ordering is shown in~\cref{fig:energy_gaps_pbesol}.

\subsection{r$^2$SCAN}
\begin{figure*}[htbp]
    \centering
    \includegraphics[width=0.98\textwidth]{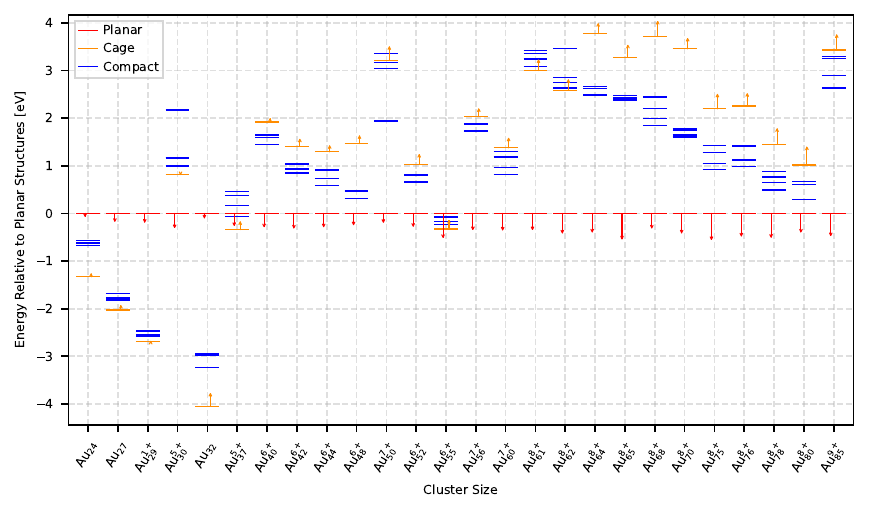}
    \caption{Total r$^2$SCAN energies for the investigated neutral and ionized gold clusters. Shown is always the smallest ionization level that gives a cage or flake as its ground state. Shown are also the vibrational free energy differences at 300 K with respect to the free energies of the lowest compact structure. Downward arrows indicate that entropy effects stabilize planar structures. 
    }
    \label{fig:energy_gaps}
\end{figure*}

r$^2$SCAN is a non-empirical, numerically regularized meta-GGA in the SCAN family that improves numerical stability while satisfying the most important known exact constraints for meta-GGA XC functionals. It builds on earlier regularization efforts rSCAN~\cite{bartok2019regularized} that were introduced to mitigate SCAN's~\cite{sun2015strongly} numerical sensitivity, while SCAN itself was originally designed as a broadly constrained meta-GGA. Within the r$^2$SCAN description, cage geometries are obtained as ground states for the neutral clusters $Au_{24}$, $Au_{27}$, and $Au_{32}$. Upon ionization, cages become the ground state for $Au_{29}^{+1}$, $Au_{37}^{+5}$, and $Au_{55}^{+6}$, whereas for all other investigated clusters, planar flakes are stabilized as ground states at the corresponding ionization levels. Overall, the ionization level required to favor planar flakes in r$^2$SCAN lies between those predicted by PBE and by PBEsol+MBD, i.e., higher than in PBE but lower than in PBEsol+MBD. For planar flake ground states, the energetic gaps to the first metastable isomer are as follows:
$Au_{61}^{+8}$, $Au_{62}^{+8}$, $Au_{64}^{+8}$, $Au_{65}^{+8}$ and $Au_{85}^{+9}$ have the gaps between $2$-$3$ eV; $Au_{40}^{+6}$, $Au_{50}^{+7}$, $Au_{56}^{+7}$, $Au_{68}^{+8}$ and $Au_{70}^{+8}$ have the gaps between $1$-$2$ eV; all remaining planar ground states have gaps of less than $1$ eV to the closest metastable structure. In the~\cref{tab:compact_symmetries}, the ionization levels required to stabilize non-compact motifs as ground states are listed. The resulting energetic ordering is shown in~\cref{fig:energy_gaps}.

\subsection{Density of states}
The compact structures have a very high density of states, i.e., there exist many nearly degenerate structures.
For each ionization level, we showed only the four lowest-energy compact structures. In general, there is a substantial energetic gap between the planar global minimum and the lowest compact cluster as seen in ~\cref{fig:energy_gaps_pbe,fig:energy_gaps_pbesol,fig:energy_gaps}. So even if there was a slightly lower compact structure, it is very unlikely that it would be lower than the planar structures.

\subsection{Temperature effects}
We investigated the effects of temperature on the stability of planar geometries by calculating the vibrational free energy at finite temperature. The results show that temperature effects further stabilize the planar geometries relative to their compact and cage counterparts. \Cref{fig:energy_gaps} shows the free energies with respect to planar geometries. It can be seen that taking into account the temperature effects changes the global minimum of $Au_{37}$ and $Au_{55}$ from a cage to a planar configuration. In addition, the gaps between planar geometries and the first metastable geometries get larger. The vibrational free energies for $Au_{50}$ to $Au_{85}$ are calculated by charge-corrected NequIP and the rest are calculated by DFT with the r$^2$SCAN XC functional at 300k. 

\subsection{Building patterns} 
\begin{figure*}[htbp]
    \centering
    \includegraphics[width=\textwidth]{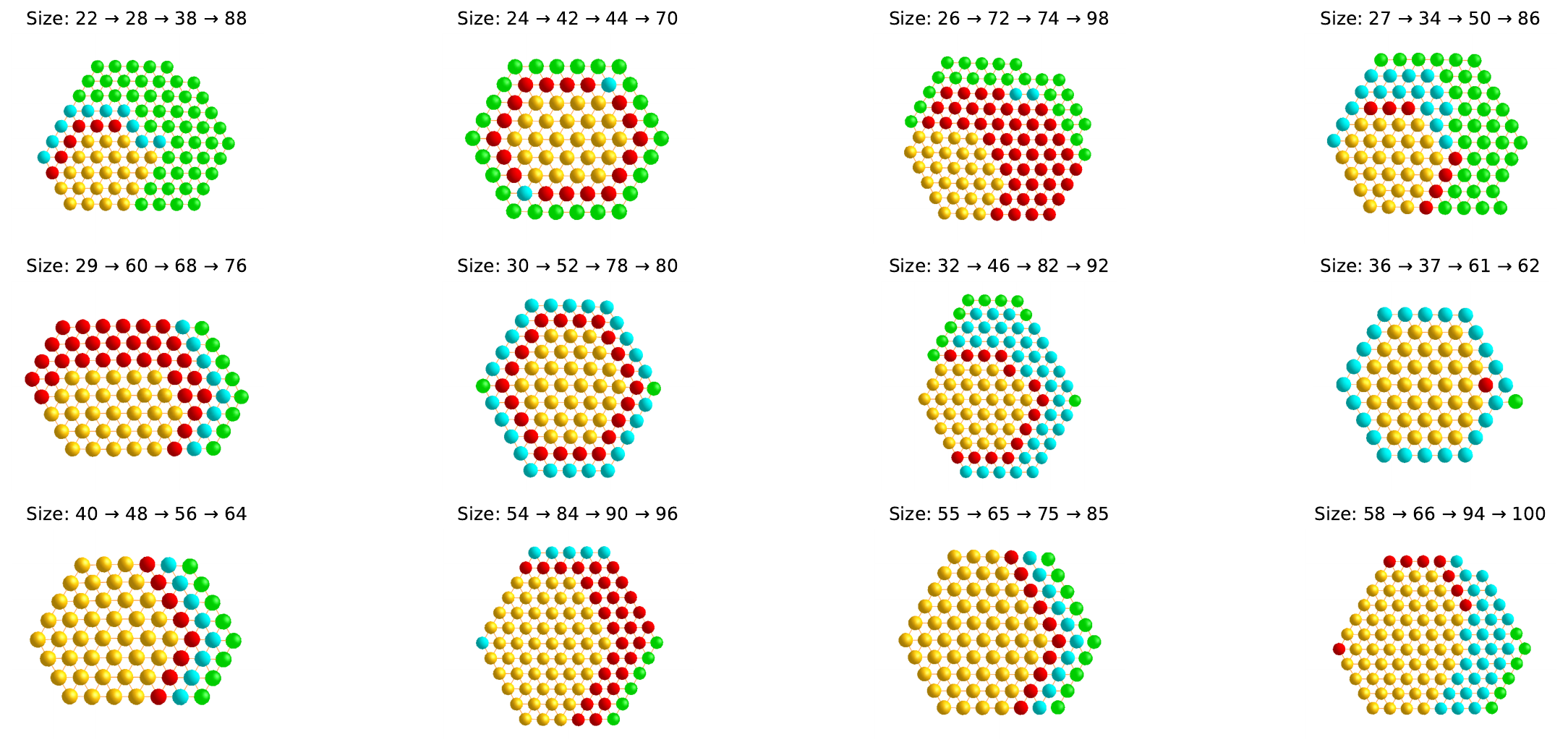}
    \caption{Planar configurations of minimally ionized gold clusters for all investigated sizes. The atoms are always arranged in a regular hexagonal close-packed pattern. The progression from one size to the next is obtained by adding to the atoms shown in gold the atoms highlighted in red, then blue and finally green.  
     }
    \label{fig:combined_planar}
\end{figure*}
To better understand the planar structural patterns, we analyzed their building patterns across multiple cluster sizes. The growth is not always achieved by completing full shells; in many cases, atoms are added selectively along edges or specific directions, leading to anisotropic yet regular structures as shown in~\cref{fig:combined_planar}. For example, the transition from 55 to 85 atoms involves additions that extend particular edges rather than forming uniform shells. Even though the required minimum ionization level for obtaining planar structures varies for different XC functionals, the lowest energy planar structures of all XC functionals have the same overall structure.
The different ionization levels and XC functionals just lead to barely visible structural relaxations. 

\subsection{Symmetries}
\begin{figure}[htbp]
    \centering
    \includegraphics[width=\columnwidth]{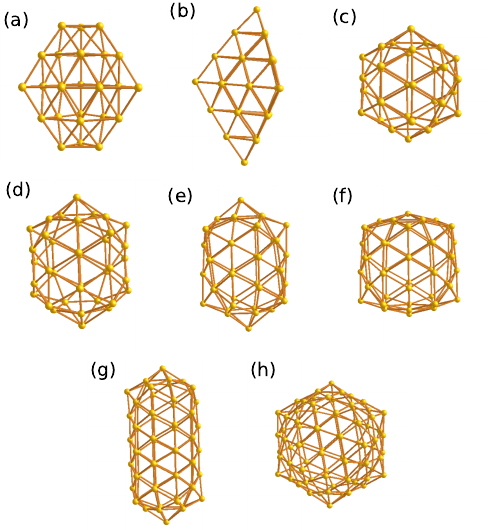}
    \caption{Subfigure (a)-(h) show the highly symmetric cage structures for $Au_{27}$, $Au_{29}$, $Au_{32}$, $Au_{37}$, $Au_{42}$, $Au_{50}$, $Au_{62}$ and $Au_{78}$ respectively.}
    \label{fig:cages}
\end{figure}
In \cref{fig:cages} we highlight high-symmetry cage structures, $Au_{32}$ ($I_h$), $Au_{27,29}$ ($D_{3h}$), $Au_{37}$ ($D_{5h}$), $Au_{42,62}$ ($D_{5d}$), $Au_{50}$ ($D_{6d}$) and $Au_{78}$ ($T_{h}$). The assigned point-group symmetries, however, are XC functional dependent. For PBE and r$^2$SCAN, these cages largely retain their high symmetry character; the only notable exception is $Au_{27}$, which relaxes from $D_{3h}$ (PBE) to $C_s$ (r$^2$SCAN). With PBEsol+MBD, $Au_{32,37,42}$ clusters relax into strongly distorted cages and lose their symmetries. Intermediate symmetries are observed in: $Au_{22,24,26}$ ($C_{2v}$), $Au_{28}$ ($C_{3v}$) and $Au_{34,75}$ ($D_{2}$). The rest of the cages, however, adopt low symmetry and are classified as $C_{s}$, $C_{2}$ and $C_{1}$. \Cref{tab:compact_symmetries} shows for the clusters that we calculated them with three XC functionals(with PBE XC functional we have calculated more clusters than the other two), the ionization level at which non-compact structures (planar or cage) become energetically competitive, together with the point-group symmetry of the corresponding compact and cage at that ionization level for r$^2$SCAN, PBEsol+MBD and PBE XC functionals. Taken together, these results demonstrate that while cage motifs frequently adopt highly symmetric forms, the compact structures favor distorted, low-symmetry configurations. Being trained on bulk materials, MACE gave only compact structures, but no low-energy cages. Finally, we note that the point-group symmetries reported here arise from the structures relaxed with a standard finite electronic smearing (Gaussian broadening) of 0.01 eV. In the zero-electronic-temperature limit, degenerate states at the Fermi level drive small Jahn-Teller distortions that lift the degeneracy by reducing the symmetry, and slightly lower the total energy. The distortions are, however, so small that they can hardly be seen by eye. In the resulting structures, all electronic levels are completely filled with integer occupation numbers, so the rule that determines the structure of small clusters, namely that clusters adopt shapes such that electronic shells can be filled completely~\cite{fisicaro}, is also applicable to larger clusters.
\begin{table}[htbp]
    \centering
    \small
    \caption{For each cluster size, the table reports for the r$^2$SCAN, PBEsol+MBD, and PBE XC functionals the ionization level $Q$ at which the lowest-energy non-compact motif (planar or cage) first becomes energetically competitive with the compact minimum, together with the point-group symmetry of the corresponding lowest-energy compact and cage structures at that $Q$.}
    \label{tab:compact_symmetries}
    \begin{tabular}{|c|ccc|ccc|ccc|}
        \hline
        \multirow{2}{*}{\parbox[c][10.5ex][c]{1.2em}{\centering\rotatebox{90}{Cluster}}}
            & \multicolumn{3}{c|}{PBE} 
            & \multicolumn{3}{c|}{PBEsol+MBD} 
            & \multicolumn{3}{c|}{r$^2$SCAN} \\
            \cline{2-10}
        & \rotatebox{90}{$Q$} & \rotatebox{90}{compact  } & \rotatebox{90}{cage}
        & \rotatebox{90}{$Q$} & \rotatebox{90}{compact  } & \rotatebox{90}{cage}
        & \rotatebox{90}{$Q$} & \rotatebox{90}{compact  } & \rotatebox{90}{cage} \\
         \hline
        $Au_{24}$ & 0  & $C_{s}$   & $C_{2v}$ & 8  & $C_{2v}$  & $C_{2v}$ & 0  & $C_{1}$ & $C_{2v}$ \\
        $Au_{27}$ & 0  & $C_{2v}$  & $D_{3h}$ & 7  & $C_{s}$   & $D_{3h}$  & 0  & $C_{s}$ & $C_{s}$ \\
        $Au_{29}$ & 1  & $C_{3v}$  & $D_{3h}$ & 7  & $C_{s}$   & $D_{3h}$  & 1  & $C_{1}$ & $D_{3h}$ \\
        $Au_{30}$ & 0  & $C_{1}$   & $C_{1}$ & 7  & $C_{s}$   & $C_{1}$  & 5  & $C_{1}$ & $C_{1}$ \\
        $Au_{32}$ & 0  & $C_{1}$   & $I_{h}$ & 7  & $C_{3}$   & $C_{1}$  & 0  & $C_{s}$ & $I_{h}$ \\
        $Au_{37}$ & 5  & $C_{1}$   & $D_{5h}$ & 8  & $C_{1}$   & $C_{2}$  & 5  & $C_{s}$ & $D_{5h}$ \\
        $Au_{40}$ & 4  & $C_{1}$   & $C_{1}$ & 8  & $C_{1}$   & $C_{1}$  & 6  & $C_{1}$ & $C_{s}$ \\
        $Au_{42}$ & 4  & $C_{1}$   & $D_{5d}$ & 9  & $C_{1}$   & $C_{1}$  & 6  & $C_{1}$ & $D_{5d}$ \\
        $Au_{44}$ & 6  & $C_{s}$   & $C_{1}$ & 9  & $C_{1}$   & $C_{1}$  & 6  & $C_{s}$ & $C_{1}$ \\
        $Au_{48}$ & 0  & $C_{s}$   & $C_{2}$ & 9  & $C_{s}$   & $C_{1}$  & 6  & $C_{s}$ & $C_{2}$ \\
        $Au_{50}$ & 0  & $C_{1}$   & $D_{6d}$ & 9  & $C_{1}$   & $D_{6d}$  & 7  & $C_{3}$ & $D_{6d}$ \\
        $Au_{52}$ & 2  & $C_{1}$   & $C_{1}$ & 9  & $C_{2v}$  & $C_{1}$ & 6  & $C_{1}$ & $C_{1}$ \\
        $Au_{55}$ & 3  & $C_{1}$   & $C_{1}$ & 10 & $C_{s}$   & $C_{1}$  & 6  & $C_{1}$ & $C_{1}$ \\
        $Au_{56}$ & 4  & $C_{1}$   & $C_{1}$ & 10 & $C_{1}$   & $C_{1}$  & 7  & $C_{1}$ & $C_{1}$ \\
        $Au_{60}$ & 6  & $C_{2}$   & $C_{2}$ & 10 & $C_{1}$   & $C_{2}$  & 7  & $C_{1}$ & $C_{2}$ \\
        $Au_{61}$ & 7  & $C_{3}$   & $C_{1}$ & 10 & $C_{1}$   & $C_{1}$  & 8  & $C_{1}$ & $C_{1}$ \\
        $Au_{62}$ & 6  & $C_{3}$   & $D_{5d}$ & 10 & $C_{3}$   & $D_{5d}$  & 8  & $C_{3}$ & $D_{5d}$ \\
        $Au_{64}$ & 6  & $C_{2}$   & $C_{1}$ & 10 & $C_{1}$   & $C_{1}$  & 8  & $C_{1}$ & $C_{1}$ \\
        $Au_{65}$ & 7  & $C_{1}$   & $C_{1}$ & 10 & $C_{1}$   & $C_{1}$  & 8  & $C_{1}$ & $C_{1}$ \\
        $Au_{68}$ & 6  & $C_{1}$   & $C_{1}$ & 10 & $C_{2}$   & $C_{1}$  & 8  & $D_{2}$ & $C_{1}$ \\
        $Au_{70}$ & 6  & $C_{1}$   & $C_{1}$ & 10 & $C_{1}$   & $C_{1}$  & 8  & $C_{1}$ & $C_{1}$ \\
        $Au_{75}$ & 7  & $C_{2}$   & $C_{2}$ & 11 & $C_{1}$   & $D_{2}$  & 8  & $C_{2}$ & $C_{2}$ \\
        $Au_{76}$ & 6  & $C_{1}$   & $C_{1}$ & 11 & $C_{1}$   & $C_{1}$  & 8  & $C_{1}$ & $C_{1}$ \\
        $Au_{78}$ & 6  & $C_{1}$   & $T_{h}$ & 11 & $C_{1}$   & $T_{h}$  & 8  & $C_{1}$ & $T_{h}$ \\
        $Au_{80}$ & 6  & $C_{1}$   & $C_{1}$ & 11 & $C_{1}$   & $C_{2}$  & 8  & $C_{1}$ & $C_{2}$ \\
        $Au_{85}$ & 7  & $C_{1}$   & $C_{1}$ & 12 & $C_{1}$   & $C_{1}$  & 9  & $C_{1}$ & $C_{1}$ \\
        \hline
    \end{tabular}
\end{table}

\section{Conclusions}
In this study, we showed that ionization can transform the ground states of medium-sized clusters from compact bulk-like structures to planar structures consisting of hexagonal honeycomb patterns.
These planar structures are considerably lower in energy than ionized compact structures of the same size. 
Weaker ionizations can also lead to cage structures.
The predicted degree of ionization that is necessary to induce cage or planar ground states, unfortunately, depends on the XC functional that is used. We found that the PBE XC functional requires the smallest amount of ionization to induce planar ground states. The r$^2$SCAN requires a somewhat 
larger ionization and PBEsol+MBD requires the largest amount. The r$^2$SCAN XC functional, which also includes dispersion interactions,  would probably be considered by most researchers as the most accurate XC functional. It yields results in closest agreement with experiment, whereas PBEsol+MBD predicts compact ground states for all neutral clusters.
For some sizes such as $Au_{24}$, $Au_{27}$ and $Au_{32}$ there are however experimental indications that the ground states 
are cages~\cite{Au24,Au27,Au32} . This is also predicted by the PBE and r$^2$SCAN XC functionals. So we also consider in this context the r$^2$SCAN results as the most reliable results. The ionization level given in~\cref{tab:compact_symmetries} should 
therefore best match an experimental verification of our results.
In spite of the disagreement concerning the minimum ionization level necessary to obtain planar structures, the 
planar structures adopted by the different XC functionals agree very well.


\section*{Conflicts of interest}
There are no conflicts to declare.

\section*{Data availability}
The coordinate files for all structures reported in this paper are deposited and are provided in the accompanying zipped archive. Further information regarding the data can be obtained by contacting the corresponding author via email.
\section*{Acknowledgements}
Computational resources were provided by the Swiss National Supercomputing Centre (CSCS) under projects s1167 and lp08, and by \href{https://scicore.unibas.ch/}{sciCORE}, the scientific computing center of the University of Basel. Financial support for this project was provided by SNF under grant number 200021\_191994. M.I.S thanks and appreciates the interesting discussions with M. Gubler, M. Sommer-Jorgensen and G. Fisicaro.


\bibliography{rsc} 
\bibliographystyle{rsc}
\newpage
\section*{Supplementary Information}
\renewcommand{\thefigure}{S\arabic{figure}} 
\setcounter{figure}{0}                        
\renewcommand{\thetable}{S\Roman{table}}
\setcounter{table}{0}
\subsection{Dataset and model setup}

\begin{figure}[htbp]
    \centering
    \includegraphics[width=0.45\textwidth]{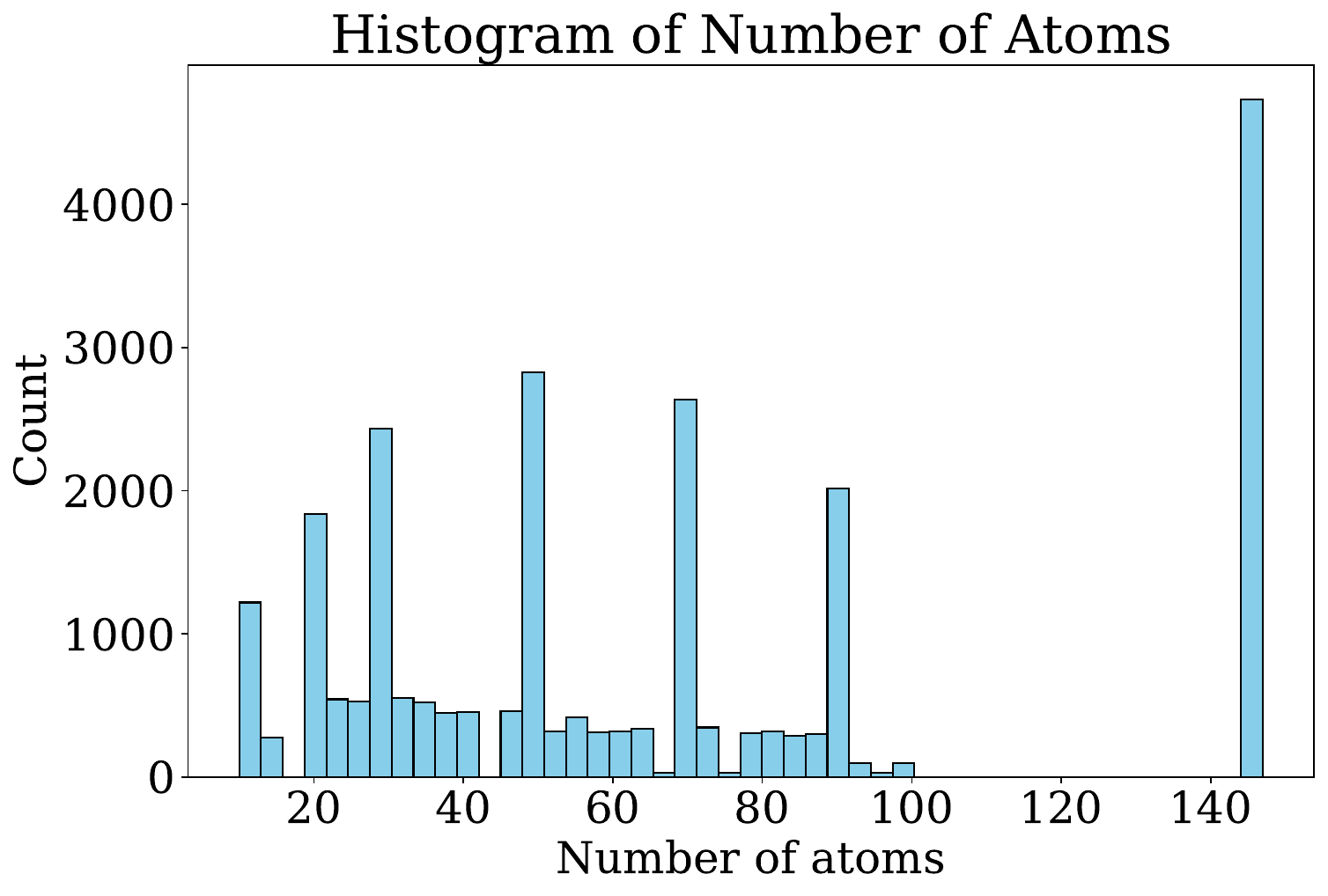}
    \caption{Histogram of the number of atoms in the training dataset.}
    \label{fig:size_distribution}
\end{figure}

\begin{figure}[htbp]
    \centering
    \includegraphics[width=0.45\textwidth]{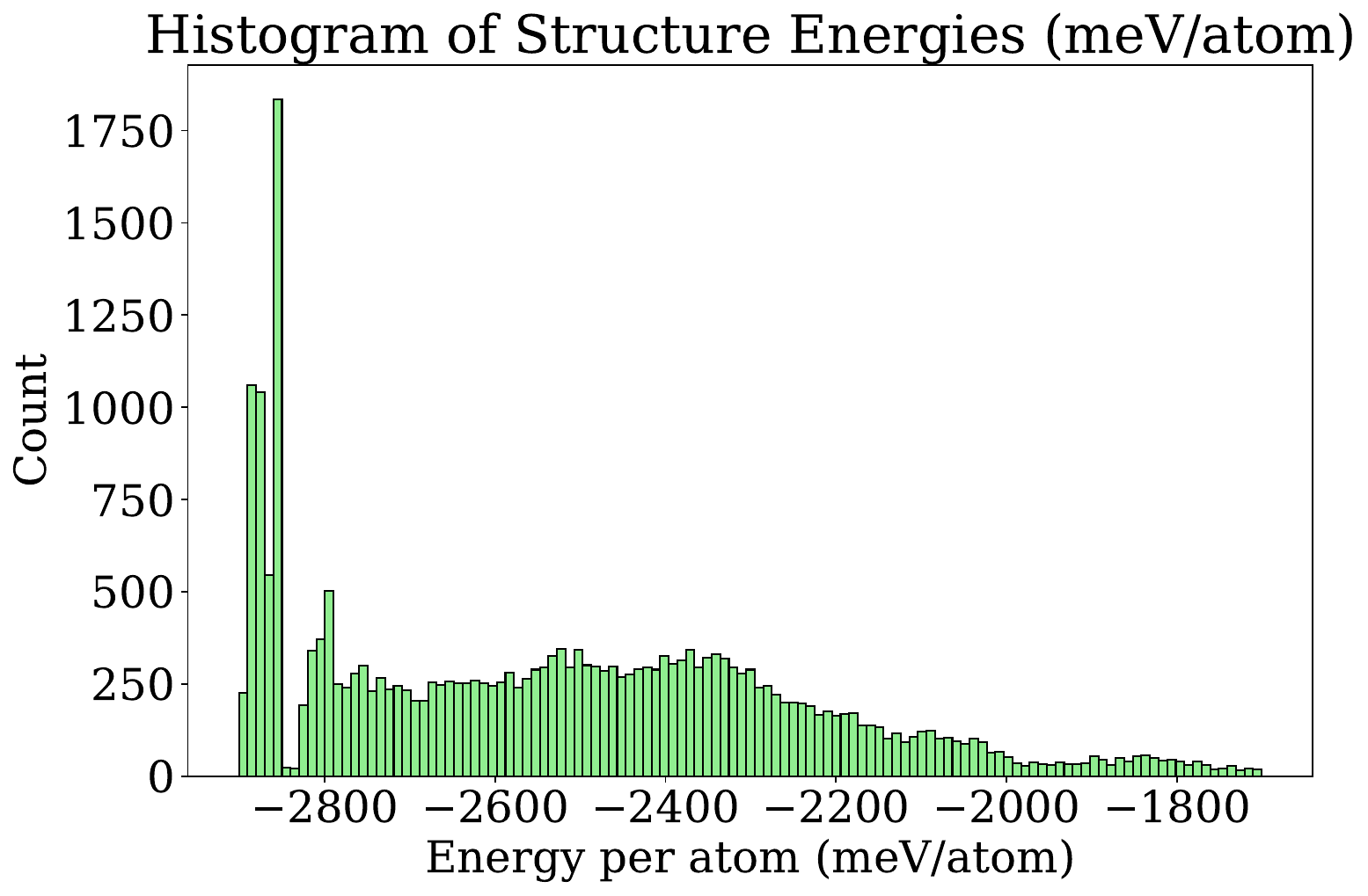}
    \caption{Energy distribution of structures in the training dataset (meV/atom).}
    \label{fig:energy_distribution}
\end{figure}

The training dataset contains a diverse set of gold clusters ranging from 10 to 140 atoms, including compact, planar, and cage structures. The DFT reference energies were computed with the Projector-Augmented Wave (PAW) method~\cite{blochl1994projector,kresse1999ultrasoft} using 11 valence electrons as implemented in the Vienna Ab initio Simulation Package (VASP)~\cite {kresse1996efficient,kresse1996efficiency} together with the PBE exchange-correlation functional~\cite{perdew1996generalized}. To remove interactions between periodic images, a vacuum of 12 \r{A} was introduced in all directions around the nanoclusters. \Cref{fig:size_distribution,fig:energy_distribution} show the distribution of cluster sizes and total energies in the training set, respectively.

The NequIP model was trained for 82 epochs using a force-matching loss that combines force and energy contributions. The network architecture, radial cutoff, number of interaction blocks, and angular-momentum channels were chosen to balance accuracy and computational efficiency.

\subsection{Performance and application}

\begin{figure}[htbp]
    \centering
    \includegraphics[width=0.45\textwidth]{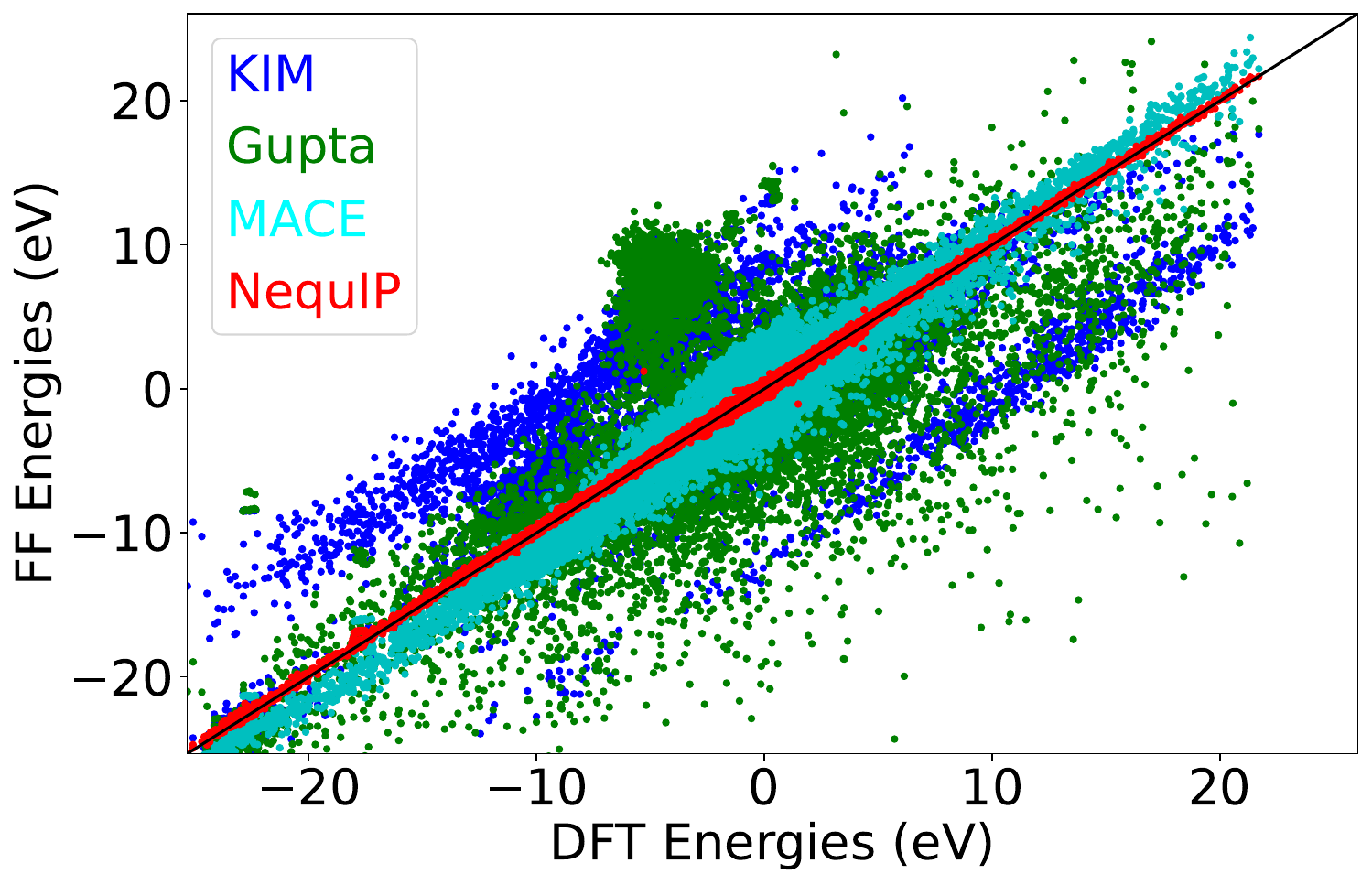}
    \caption{Correlation of total energies by EAM method ~\cite{daw1984embedded,o2018grain}, Gupta~\cite{gupta1981lattice}, MACE and NequIP models with DFT reference energies for a representative set of gold cluster configurations.}
    \label{fig:correlation_plot}
\end{figure}

The final NequIP model reproduces the DFT reference data with high accuracy for both energies and forces and is therefore suitable as a surrogate potential for global structure searches. Detailed performance metrics, including root-mean-square errors (RMSEs) for energies and forces on the training and validation sets, are summarized in~\cref{tab:train_info}. A comparison with other interaction models is shown in~\cref{fig:correlation_plot}.

\begin{table}[htbp]
    \centering
    \caption{Energy and force root-mean-square errors of the trained NequIP model on the training and validation sets.}
    \renewcommand{\arraystretch}{1.1}
    \begin{tabular}{lcc}
        \hline
        & $E_{\mathrm{RMSE}}$ (eV) & $F_{\mathrm{RMSE}}$ (eV/\AA) \\
        \hline
        Training   & 0.00541 & 0.0463 \\
        Validation & 0.00441 & 0.0473 \\
    \end{tabular}
    \label{tab:train_info}
\end{table}
\subsection{Validation of the charge-corrected NequIP potential}

\begin{figure}[htbp]
    \centering
    \includegraphics[width=0.45\textwidth]{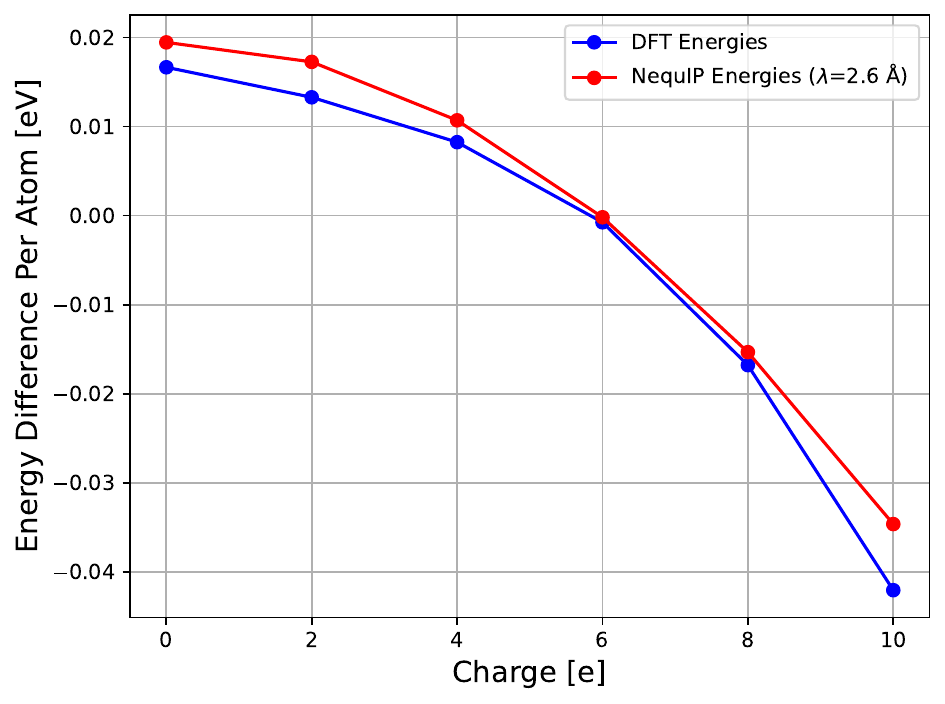}
    \caption{Energy difference per atom as a function of charge for $Au_{76}$, comparing DFT and the charge-corrected NequIP potential. Energies are referenced to the compact structure.}
    \label{fig:energy_Au76}
\end{figure}

\begin{figure}[htbp]
    \centering
    \includegraphics[width=0.45\textwidth]{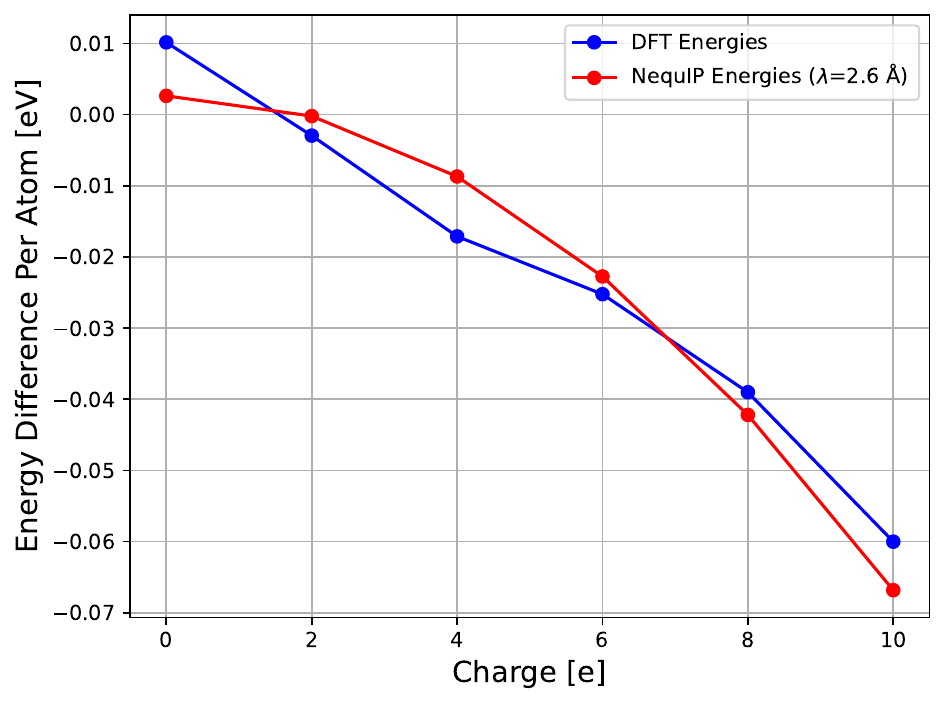}
    \caption{Energy difference per atom as a function of charge for $Au_{54}$, comparing DFT and the charge-corrected NequIP potential. Energies are referenced to the compact structure.}
    \label{fig:energy_Au54}
\end{figure}

To assess whether the electrostatic correction preserves the quality of the underlying NequIP potential, we first compared DFT and charge-corrected NequIP energies as a function of ionization level for representative clusters.~\Cref{fig:energy_Au76,fig:energy_Au54} show the energy difference per atom relative to the compact structures for $Au_{76}$ and $Au_{54}$, respectively, over the charge range $q = 0$-$10e$. In both cases, the charge-corrected NequIP curves closely track the DFT trends: the energies decrease monotonically with increasing charge. The deviations between DFT and the corrected NequIP energies remain small compared to the total stabilization with charge, indicating that the added electrostatic term does not distort the global charge dependence of the PES.

\begin{figure}[htbp]
    \centering
    \includegraphics[width=\columnwidth]{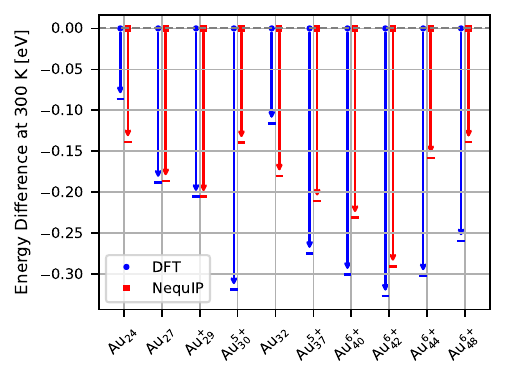}
    \caption{Vibrational free-energy differences $\Delta F_{\mathrm{vib}}(300~\mathrm{K})$ between planar and compact structures for selected clusters, comparing DFT and the charge-corrected NequIP potential. Negative values indicate that the planar structure is favoured at 300K.}
    \label{fig:fvib_planar_compact}
\end{figure}

\begin{figure}[htbp]
    \centering
    \includegraphics[width=\columnwidth]{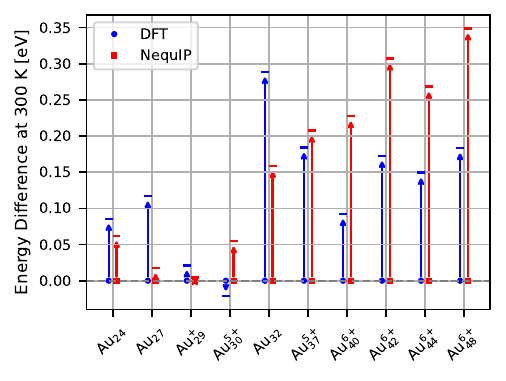}
    \caption{Vibrational free-energy differences $\Delta F_{\mathrm{vib}}(300~\mathrm{K})$ between cage and compact structures for selected clusters, comparing DFT and the charge-corrected NequIP potential. Positive values indicate that the cage structure is higher in free energy than the corresponding compact minimum at 300K.}
    \label{fig:fvib_cage_compact}
\end{figure}

Since the vibrational free energy depends on the curvature of the PES around the minimum, a comparison of vibrational free-energy differences provides an implicit but stringent test of the forces and their derivatives.~\Cref{fig:fvib_planar_compact,fig:fvib_cage_compact} show the vibrational free-energy differences at $T = 300$~K (including zero-point energy) between planar and compact structures, and between cage and compact structures, respectively. For all considered cluster sizes, the charge-corrected NequIP results closely follow the DFT values: the sign of the free-energy difference is identical in every case, the magnitude is reproduced to good accuracy, and the relative ordering of clusters is preserved. In particular, both DFT and charge-corrected NequIP predict a substantial lowering of the free energy for planar structures relative to compact ones, and a corresponding increase for cage structures relative to their compact counterparts.

Taken together, these comparisons demonstrate that the electrostatic correction preserves both the global charge dependence of the energy and the local curvature around relevant minima. The charge-corrected NequIP potential therefore provides a reliable description of energies and forces for the ionized gold clusters studied here.

\end{document}